\documentclass{andp2012}
\usepackage[english]{babel}

\usepackage[utf8]{inputenc}
\usepackage[T1]{fontenc}
\usepackage{amsmath}
\usepackage{hyperref}
\usepackage[numbers,sort&compress]{natbib}
\usepackage{array}
\newcolumntype{C}[1]{>{\centering\arraybackslash}p{#1}}
\newcolumntype{R}[1]{>{\raggedright\arraybackslash}p{#1}}
\newcolumntype{L}[1]{>{\raggedleft\arraybackslash}p{#1}}

\usepackage{tikz}
\usepackage{lipsum}

\usepackage{amsmath,epsfig,graphicx,amssymb,slashed}
\usepackage{chngcntr}
\counterwithin{figure}{subsection}
\counterwithin{table}{subsection}

\keywords{local realism, Bell test, no-signaling}
\title{Analysis of assumptions in BIG Bell Test experiments}
\author[S. So{\l}tan]{Stanis{\l}aw So{\l}tan\inst{1}}
\author[D. Dopier{\l}a]{Dawid Dopiera{\l}a\inst{1}}
\author[A. Bednorz]{Adam Bednorz\inst{1}\footnote{E-mail:~\textsf{abednorz@fuw.edu.pl}}}
\address[1]{Faculty of Physics, University of Warsaw, ul. Pasteura 5, PL02-093 Warsaw, Poland}
\shortauthors{S. So{\l}tan et al.}

\begin{abstract}
Recently, a group of experiments tested local realism with random choices prepared by humans.
These various tests were subject to additional assumptions, which lead to loopholes in the interpretations of almost all of the experiments.
Among these assumptions are fair sampling, no signaling, and faithful reproduction of a Bell-type quantum model.
We examined the data from 9 of 13 experiments and analyzed occurring anomalies in view of the above assumptions.
We conclude that further tests of local realism need better setup calibration to avoid apparent signaling or necessity of the complicated underlying quantum model.

\end{abstract}

\shortabstract
\begin{document}
\maketitle

\section{Introduction}

Rejection of local realism models is a theoretical and experimental challenge.
The original argument for local realism by Einstein, Podolsky and Rosen \cite{epr}, has later been turned into a form of testable inequalities governing measurement outcomes in such models by Bell and others \cite{bell,chsh,eber}. 
The test assumes that two (or more) observers are separated and choose what to measure. Due to separation, a principle of special relativity applies that no information can be communicated outside the forward light cone. This is no-signaling and it cannot be deduced from the quantum field theory alone \cite{wight,rel}. No-signaling is natural in large setups with relativistic scales. 
A weaker definition of no-signaling fitting also the non-relativistic setups would be the absence of any interactions between parties that could transfer information before the measurement is completed. 
The opposite, signaling, would mean that such transfer somehow occurred.

Local realism means the existence of a joint (positive) probability for all choices to have specific outcomes, under the assumption of locality and predetermination or local causality and free choice\cite{assum}. Here locality implies that the outcome can depend only on the local choice (i.e. the choice and the measurement are made by the same party). Locality implies no-signaling but not vice versa.
Correlations satisfy certain inequalities if local realism is assumed. 
Their violation indicates a violation of local realism or conflict with no-signaling assumption.
 Bell test is stronger 
than steering where one assumes quantum representation of observables at one of parties \cite{eprst1,eprst2,eprst3}.

The tests of local realism were realized in the past usually with photons \cite{belx1a,belx1b,belx1c,belx1g,belx2a,belx2b}.
However, no specific system is required, it only has to fit the simple quantum few-state approximation, which resulted in setups across nearly 
all branches of physics \cite{belx1d,belx1e,belx1f,belx1h}, see the review \cite{genov}. Unfortunately, among various problems, the most significant appeared lack of sufficient  distance (locality loophole), imperfect detection (detection loophole, e.g. a fraction of particles are lost) \cite{loop1,loop2,graft} and predetermined (often 
fixed) choices (random or free choice hypothesis). Loopholes allow for a local realistic model \cite{loopmod1,loopmod2}.

In the recent Bell tests performed in Delft \cite{hensen}, NIST \cite{nist}, Vienna \cite{vien} and Munich \cite{munch}, claimed as loophole-free,  violation of local realism is claimed with high confidence level (assuming local realism, the probability of the data is 4\% in \cite{hensen}, $\sim 10^{-7}$ in \cite{nist}, $\sim 10^{-31}$ in \cite{vien} and $\sim 10^{-9}$ in \cite{munch}). However, all these experiments show also some moderate anomalies, that need
either signaling or a complicated underlying quantum model to explain \cite{ab17}. Moreover, the choice in these experiments was randomly generated by a machine, which 
cannot exclude some conspiracy models, where the choice is controlled by the other party. To rule out this possibility, one should use human-generated 
choices. In principle, to prevent locality loophole, the choices should be quicker than the time of communication between observers, which would need Moon-Earth distance \cite{moon}. A weaker challenge has been undertaken in the BIG Bell test (BBT) \cite{bbt}, where the human choices were collected all over the Earth, with a too long timescale to close this locality loophole in the strict sense. Nevertheless, trusting that the remote observers have no
access to the other party's choices, BBT can indeed test local realism. In the test, the stream of bits (either $0$ or $1$) was generated by Bellsters (self-selected people) and used to control choices of separate parties, which performed a test of local realism. In 13 various tests, local realism has been violated, but
usually with additional assumptions. They have often assumed ideal quantum Bell states and measurement angles. The goal of the hereby analysis is to examine the data for these assumptions. We requested the data from all 13 experiments
and obtained them from 9. Below we present the analysis of each of these 9 experiments, discussing anomalies given the type of experiment and known
technical features. Experiment 3 has been analyzed separately \cite{liu} but without any discussion of signaling and deviations from the underlying quantum model. The findings are later summarized in the discussion, with improvement recommendations for future tests of local realism.

\section{Methodology and main tested hypotheses}

Local realism means existence of the joint positive probability $\tilde{p}(A_0,A_1,B_0,B_1)$ for $A_a$ being the outcome measured by the observer $A$
who chose $a$ (e.g. the measurement basis). The locality excludes dependence on the remote choice, i.e. hidden random variables $A_{ab}$ are not allowed.
It generalizes to three or more observers by adding $C_c$ etc. to the set of random variables. In an experimental test, one has to assume
 absence of communication, i.e. information about the setting choice of the party $A$ cannot reach the party $B$ (and $C$, $D$. etc. in the multipartite case) before completion of its measurement. This assumption cannot be verified, only falsified. 
The measurable probability is a marginal of $\tilde{p}$ for two or three parties, i.e.
\begin{eqnarray}
&&\tilde{p}(AB|XY)=\sum_{A_{\bar{X}},B_{\bar{Y}}}\nonumber\\
&&\tilde{p}(A=A_X,B=B_Y,A_{\bar{X}},B_{\bar{Y}}),\\
&&\tilde{p}(ABC|XYZ)=\sum_{A_{\bar{X}},B_{\bar{Y}},C_{\bar{Z}}}\nonumber\\
&&\tilde{p}(A=A_X,B=B_Y,C=C_Z,A_{\bar{X}},B_{\bar{Y}},C_{\bar{Z}}),\nonumber
\end{eqnarray}
where $X,Y,Z=0,1$ are the chosen settings while $\bar{X}=1-X$, etc., denote complementary (not measurable on the same state).
Within the received data, we test the statistical independence of measurements by one observer from the choice of measurement basis by the other, i.e. no-signaling, being the null hypothesis, i.e.
\begin{equation}
\tilde{p}(\ast B|0Y)=\tilde{p}(\ast B|1Y),\;
\tilde{p}(A\ast|X0)=\tilde{p}(A\ast|X1)
\end{equation}
where $\ast$ means ignoring (summing over) that outcome. In an ideal test, we could simply check the above inequality, as done e.g. in \cite{ab17}. Unfortunately,  BBT experiments drop some events, and the probability to get outcome $0$ or $1$ is of finite efficiency $\eta$ not equal $100\%$. It can depend on various factors, but the most important is setting dependence. We test the assumption that the measured probability $p$
is related to $\tilde{p}$ by an efficiency dependent only locally on the setting, i.e.
\begin{equation}
p(AB|XY)=\tilde{p}(AB|XY)\eta_a(X)\eta_b(Y)\label{nnpp}
\end{equation}
with efficiencies $\eta<1$. Note that additional possible efficiency dependent locally on the outcome is already included and does not alter the no-signaling hypothesis. However, a combined dependence of $\eta(A,X)$ is beyond the scope of this analysis as it gives too many free parameters. The only exception is 
when testing a particular quantum model, which we do for experiments 4 and 12.

\begin{figure}
\includegraphics[scale=0.7]{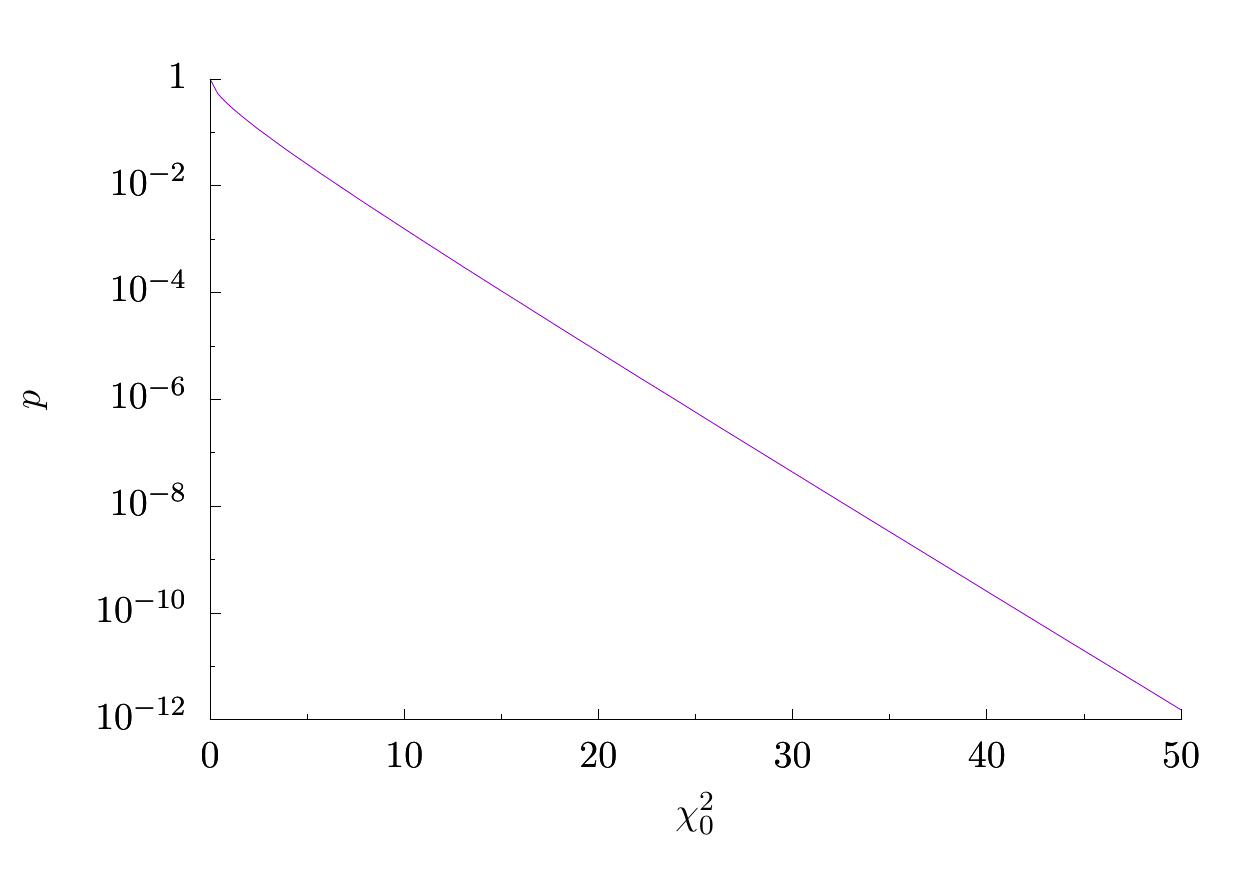}
\caption{The logarithmic dependence of the $p$-value from $\chi^2_0$.}\label{p-chi}
\end{figure}

We could first determine $\eta$ from the data and then test (\ref{nnpp}) but this is unnecessarily complicated. Instead,
we keep parameters $\eta$ as unknown but constant so the no-signaling hypothesis becomes an independence hypothesis.
Then we can employ the standard Pearson's $\chi^2$ test \cite{taylor}, with the independence being our null hypothesis. Let  $O_{ab}$ be the number of measurements for which the two observers have each assigned one parameter $a$ and $b$ respectively, with two options ($a,b=0,1$). These parameters could be the outcome of one observer and the experimental setup of the other. The numbers for each of these measurements considered separately are $O_b=\sum_aO_{ab}$ and $O_a=\sum_bO_{ab}$, $N=\sum_{ab}O_{ab}=\sum_aO_a=\sum_bO_b$ is the total count of the registered outcomes. If the probability of the measurement
 is statistically independent with respect to $a$ and $b$ then $O_{ab}$ is expected to be equal to $E_{ab}=Np_ap_b$ with $p_a=O_a/N$ and $p_b=O_b/N$ being the probabilities of specific events, which is the statement of the null hypothesis. The $\chi^2=\sum_{ab}(O_{ab}-E_{ab})^2/E_{ab}$ value is expected to be close to $0$ in order for the null hypothesis (the independence) to hold. More specifically, for $\chi^2$ from a given experiment one could obtain the probability ($p$-value) as the upper tail region of $\chi^2$ distribution for one degree of freedom above a specific value $\chi_0^2$. For large $N$, it is equal $\mathrm{erfc}(\chi_0/\sqrt{2})$ by central limit theorem, and
$p\sim \exp(-\chi^2_0/2)$ for large $\chi$, see Fig. \ref{p-chi} for the general behavior. In the case of no-signaling, we test independence of $p(A|XY)=p(A)\eta(X)\eta(Y)$ for a given $X$ and $2\times 2$ table of $A$ and $Y$, and similarly for $B$.
The obtained $p$-value has to be increased by Bonferroni correction, a.k.a. look-elsewhere-effect \cite{bonf,lee}, i.e. multiplied by the number of
possible tests (4 in the two-party case).

\section{BBT experiments}

The BBT is a group of 13 experiments. We asked for the data of all of them and received them from 9, i.e. 1,2,4,5,6,9,10,12,13.
Below, we present the analysis of each of these 9 experiments, with a short description of the setup and aim, and then an explanation of the
analyzed assumptions and found anomalies, according to the above methodology.

All of the experiments discussed here used photons and most of them suffer from loopholes \cite{larsson}. Most of the experimenters applied fair sampling in their analysis, i.e. the instances in which all relevant photons that were measured are considered representative for all the instances produced, and the instances in which a photon was lost are discarded. The distances are too small to prevent all subluminal influence. Only experiment 13 is free from these loopholes.

\subsection{1: R.B. Patel et al. Quantum steering using human randomness}
 
\begin{figure}
\includegraphics[scale=0.6]{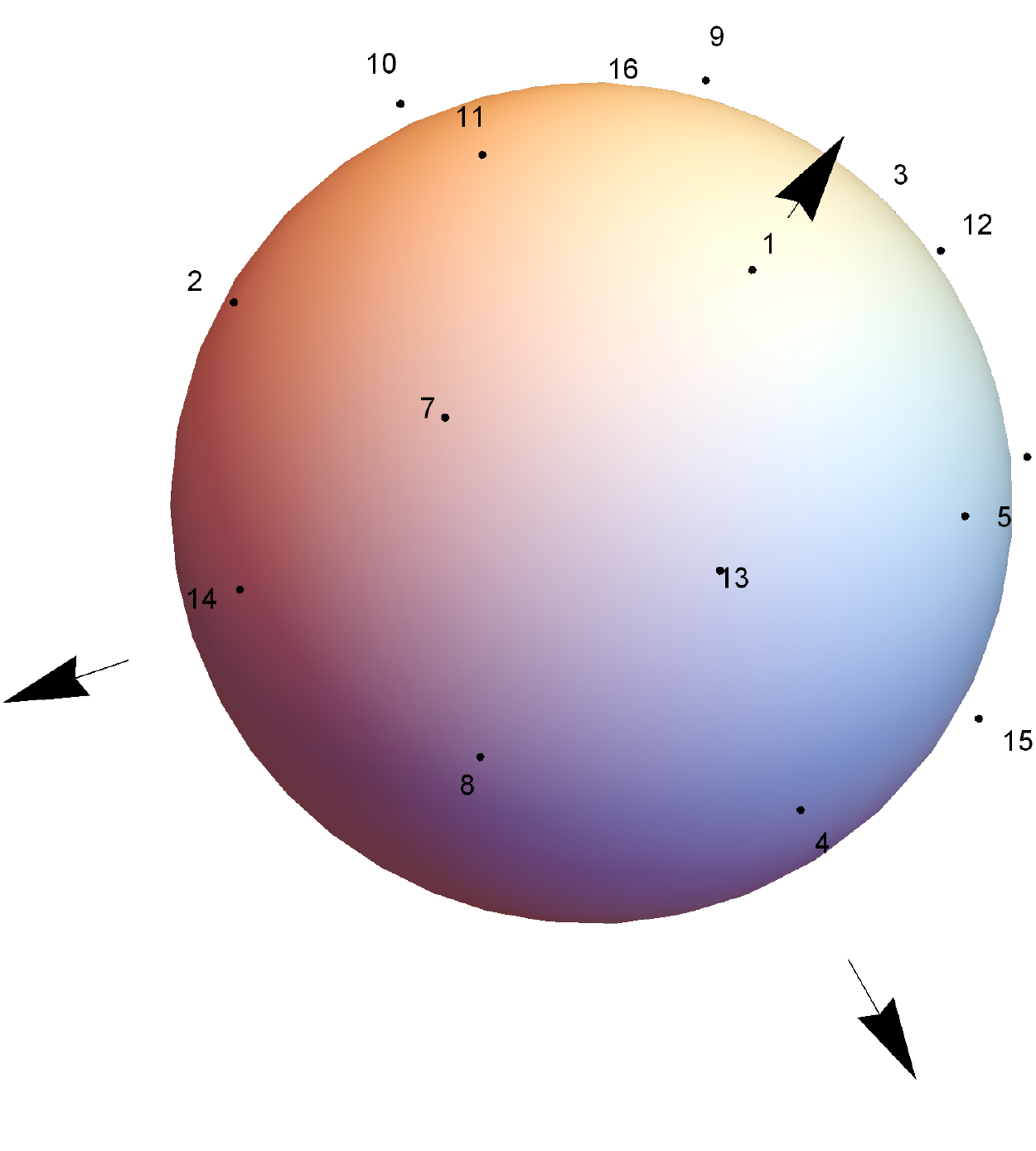}
\caption{The directions of the measured operators on the Bloch sphere}\label{blo}
\end{figure}

 The experiment tested steering on Bell state $|\psi\rangle=(|+-\rangle-|-+\rangle)/\sqrt{2}$,  measuring photons with random pairs of settings, which are the same for both measuring parties.
 Essentially the measured quantity was $\hat{A}_k=\boldsymbol n_k\cdot\hat{\boldsymbol\sigma}$, with Pauli matrices $\hat{\sigma}_1=|+\rangle\langle -|
 +|-\rangle\langle +|$, $\hat{\sigma}_2=i|-\rangle\langle +|-i|+\rangle\langle -|$, $\hat{\sigma}_3=|+\rangle\langle +|-|-\rangle\langle -|$.
 The 16 unit ($|\boldsymbol n|=1$) directions $\boldsymbol n_k$ are located at vertices and centers of sides of a dodecahedron (or centers of sides of a regular soccer ball) with $\boldsymbol n_1=(0,0,1)$ being one of the dodecahedron's vertices (see Figure \ref{blo}).
 The settings are controlled by a half-wave-plate and quarter-wave-plate. In general Bell correlations read
 \begin{equation}
 \langle \hat{A}_k\hat{B}_j\rangle=-\boldsymbol n_k\cdot\boldsymbol n_j=-\cos\phi_{kj}\label{bellcor}
 \end{equation}
 where $\phi_{kj}$ is the angle between directions $\boldsymbol n_k$ and $\boldsymbol n_j$.
Here, ideally the correlations should be $\langle \hat{A}_k\hat{B}_k\rangle=-1$ with $A_k,B_k=\pm 1$. Instead of fair sampling, applied usually to the fraction of 
 coincidences in the set of all events (including $A_k,B_k=0$), this experiment has oversampling, i.e. for a given setting, a large group of states is measured
 \emph{until} the coincidence is registered. Since the settings are the same, we could not check signaling i.e. dependence of one party's measured probability on the choice of the other. The only analysis we could do is the occurrence of a particular setting.

In Table \ref{tab1}, it is clear the choices 6 and 11 occur much often. According to authors \cite{raj} it can be attributed to
anticorrelations from Bellsters. The bits provided by humans are usually not fully random, and this increased occurrence may be the result of combinations like $0101$ and $1010$ (binary representations of 5 and 10 respectively). 
However, this should not cause any new loopholes.

\begin{table}
\begin{tabular}{|c|c|}
\hline
setting & occurrence\\
\hline
1& 1549\\
2& 1478\\
3& 1744\\
4& 716\\
5& 1448\\
6& 3947\\
7& 988\\
8& 736\\
&\\
\hline
\end{tabular}
\begin{tabular}{|c|c|}
\hline
setting & occurrence\\
\hline
9& 922\\
10& 1529\\
11& 3575\\
12& 900\\
13& 834\\
14& 1274\\
15& 1239\\
16& 1238\\
\hline
all&  24117\\
\hline
\end{tabular}
\caption{Numbers of occurrences of particular settings}\label{tab1}
\end{table}

\subsection{2: M. Ringbauer and A. White, Quantum Correlations in Time}
\label{ex2}
 
\begin{figure}
\includegraphics[scale=0.75]{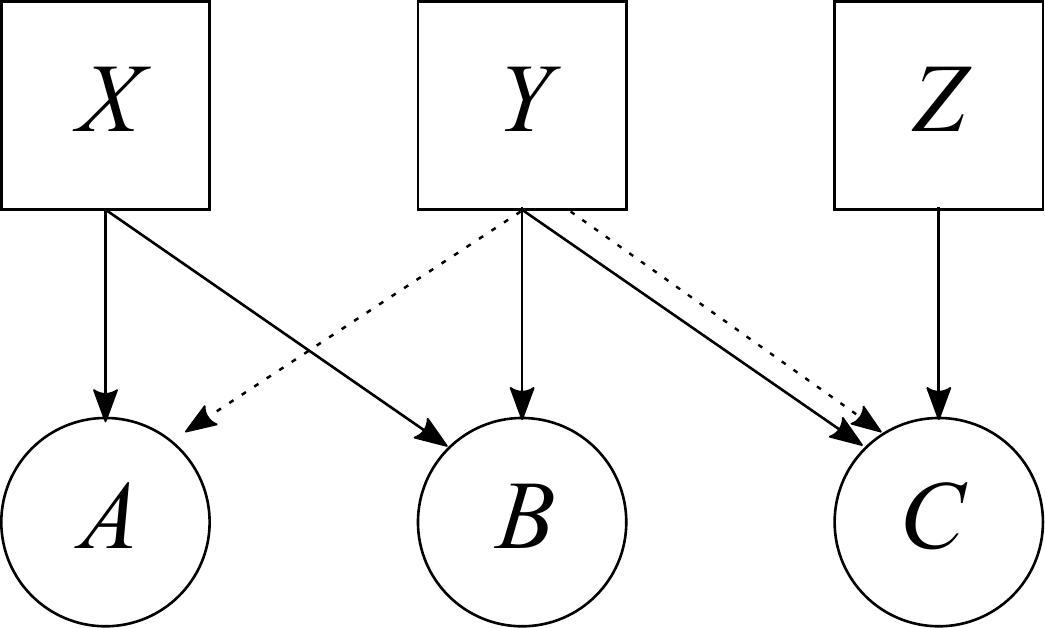}
\caption{Allowed causal order in experiment 2 (solid) and the apparent signalling (dashed).}
\label{xyzabc}
\end{figure}

 The experiment tested local realism translated into time correlations. Essentially, the setup consists of three observers $A$, $B$, $C$, each measuring $0$ or $1$ at respective choice $X$, $Y$, $Z$ (in fact $A$ measurement means a preparation of the state), made by an appropriate setting of halfwaveplates (HWP)
and quarterwaveplates (QWP). The sequence of measurement allows the causal order depicted in Fig. \ref{xyzabc} and the violation of the classical correlation assumes this order. Due to finite detection efficiency, fair sampling is assumed.

We checked if the causal order is satisfied in sense of no-signaling, i.e. if the probability cannot depend on the choice not linked causally, see Figure \ref{xyzabc}. Let us denote the probability $p(ABC|XYZ)$ of measuring $A$, $B$ and $C$ for the choices $X$, $Y$, $Z$. Because of postselection of coincidences, we assumed also
detection efficiency to depend only on the local choice but not on the outcome. It means in particular that 
\begin{eqnarray}
&&p(ABC|XYZ)=\label{nnnppp}\\
&&\tilde{p}(ABC|XYZ)\eta_a(X)\eta_b(Y)\eta_c(Z)\nonumber
\end{eqnarray}
with efficiencies $\eta<1$.
Then no-signaling implies $\tilde{p}(\ast\ast C|0YZ)\tilde{p}(\ast\ast C|1YZ)$, $\tilde{p}(\ast B\ast|XY0)=\tilde{p}(\ast B\ast|XY1)$, $\tilde{p}(A\ast\ast|XYZ)=\tilde{p}(A\ast\ast|X\ast\ast)$,
$\tilde{p}(AB\ast|XY0)=\tilde{p}(AB\ast|XY1)$. Here $\ast$ means ignoring/ discarding outcome or every choice.
The summed count for each combination of choices and outcomes is shown in Table \ref{tt1}.

\begin{table}
\begin{tabular}{|C{0.45cm}|C{0.45cm}C{0.45cm}C{0.45cm}C{0.45cm}C{0.45cm}C{0.45cm}C{0.45cm}C{0.45cm}|}
\hline
&000&001&010&011&100&101&110&111\\
\hline
000& 6& 23& 114& 16& 36& 240& 47& 5\\
001& 7& 14& 111& 21& 46& 245& 48& 6\\
010& 5& 25& 118& 21& 42& 187& 31& 8\\
011& 21& 9& 15& 131& 229& 39& 5& 25\\
100& 9& 18& 120& 13& 38& 226& 52& 2\\
101& 5& 31& 123& 30& 40& 219& 30& 7\\
110& 57& 224& 47& 8& 5& 18& 97& 31\\
111& 233& 23& 6& 44& 20& 2& 11& 110\\
\hline
\end{tabular}
\caption{Statistics of coincidences (columns $ABC$) at given choices (rows $XYZ$) of experiment 2}
\label{tt1}
\end{table}
\begin{table}
\centering{
\begin{tabular}{|c|c|c|c|c|}
\hline
&00*&01*&10*&11*\\
\hline
000& 29& 130& 276& 52\\
001& 21& 132& 291& 54\\
010& 30& 139& 229& 39\\
011& 30& 146& 268& 30\\
100& 27& 133& 264& 54\\
101& 36& 153& 259& 37\\
110& 281& 55& 23& 128\\
111& 256& 50& 22& 121\\
\hline
\end{tabular}
\begin{tabular}{|c|c|c|c|c|}
\hline
&0*0&0*1&1*0&1*1\\
\hline
000& 120 & 39 & 83 & 245\\
001& 118 & 35 & 94 & 251\\
010& 123 & 46 & 73 & 195\\
011& 36 & 140 & 234 & 64\\
100& 129 & 31 & 90 & 228\\
101& 128 & 61 & 70 & 226\\
110& 104 & 232 & 102 & 49\\
111& 239 & 67 & 31 & 112\\
\hline
\end{tabular}
\begin{tabular}{|c|c|c|c|c|}
\hline
&*00&*01&*10&*11\\
\hline
000& 42 & 263 & 161 & 21\\
001& 53 & 259 & 159 & 27\\
010& 47& 212& 149 & 29\\
011& 250 & 48 & 20 & 156\\
100& 47 & 244 & 172 & 15\\
101& 45 & 250 & 153 & 37\\
110& 62 & 242 & 144 & 39\\
111& 253 & 25 & 17 & 154\\
\hline
\end{tabular}}
\caption{Counts of Table \ref{tt1} with one of outcomes ignored (marked by $\ast$)}
\label{tt2}
\end{table}
\begin{table}
\centering{
\begin{tabular}{|c|c|c|}
\hline
&0**&1**\\
\hline
000& 159 & 328\\
001& 153 & 345\\
010& 169 & 268\\
011& 176 & 298\\
100& 160 & 318\\
101& 189 & 296\\
110& 336 & 151 \\
111& 306 & 143\\
\hline
\end{tabular}
\begin{tabular}{|c|c|c|}
\hline
&*0*&*1*\\
\hline
000& 305 & 182\\
001& 312 & 186\\
010& 259 & 178\\
011& 298 & 176\\
100& 291 & 187\\
101& 295 & 190\\
110& 304 & 183\\
111& 278 & 171\\
\hline
\end{tabular}
\begin{tabular}{|c|c|c|c|}
\hline
&**0&**1 & total\\
\hline
000& 203 & 284 & 487\\
001& 212 & 286 & 498\\
010& 196 & 241 & 437\\
011& 270 & 204 & 474\\
100& 219 & 259 & 478\\
101& 198 & 287 & 485\\
110& 206 & 281 & 487\\
111& 270 & 179 & 449\\
\hline
\end{tabular}}
\caption{Counts of Table \ref{tt1} with two outcomes ignored (marked by $\ast$) and total numbers of coincidences}
\label{tt3}
\end{table}
\begin{table}
\centering{
\begin{tabular}{|c|c|c|c|c|}
\hline
& 00 & 01 & 10 & 11\\
\hline
X and A& 0.07 & 7.3 & 85.4 & 88.9 \\
X and B& 0.3 & 0.3 & 0.9 & 0.08 \\
Y and B& 1.09 & 0.004 & 0.2 & 0.1 \\
Y and C& 0.9 & 20.1 & 1.2 & 34.7 \\
Z and C& 0.07 & 13.3 & 2.4 & 29.7 \\
\hline
\end{tabular}}
\caption{$\chi^2$ values calculated for causally linked choices and outcomes. It is expected to fail here. Each test is performed for other choices fixed; their values are given in the upper row in alphabetic order (XY, YZ or XZ).}
\label{tt4}
\end{table}
\begin{table}
\centering{
\begin{tabular}{|c|c|c|c|c|}
\hline
& 00 & 01 & 10 & 11\\
\hline
X and C& 1.6 & 0.3 & 0.6 & 0.9 \\
Y and A& 3.6 & 4.4 & 121.8 & 79.7 \\
Z and A& 0.4 & 0.2 & 3.1 & 0.07 \\
Z and B& 5.2$\times 10^{-5}$ & 1.2 & 0.0002 & 0.02 \\
\hline
\end{tabular}}
\caption{$\chi^2$ values calculated for causally independent choices and outcomes. Each test is performed for other choices fixed; their values are given in the upper row in alphabetic order (XY, YZ or XZ).}
\label{tt4a}
\end{table}

No-signaling  means that $N(A\ast\ast|XYZ)$ is independent of $YZ$. However, the differences (see  Table \ref{tt4a} ) are so large (e.g. $\chi^2\sim 121$ for $XZ=10$ while comparing each $Y$ and $A$) that without question they must 
originate from a systematic effect. According to the experiment authors \cite{martin}, this apparent signaling is due to the special Bob's measurements which
suppresses the rate of the measurement of $A=0$ except the case $X=Y=1$ when $A=1$ is suppressed.
The  apparent signaling $Y\to C$ is visible in $N(\ast\ast C|1Y1)$ with $\chi^2\sim 34$ ($p\sim 10^{-7}$). It is allowed by the causality flow. However, this also may be a systematic effect revealed in
$YZ=11$ in two last rows of Tables \ref{tt2}. 
The other no-signaling tests are passed within the acceptable certainty level, but also the statistics is relatively small.

The dependence of the coincidences on the setting is unquestionable. The possible reason is angle-dependent deflection at HWP or QWP which
changes the cross section between the photon wavepacket and the fiber. It could be also back-signaling due to small distances compared to time photons need to pass their routes, but we cannot check it here. Nevertheless, the effect is so large that it should be possible to run a diagnostic test 
to confirm the cause in the future.

\subsection{4: B. Liu et al., Violation of a Bell Inequality using Entangled Photons and Human Random Numbers}
\label{sec4}
 
 The experiment tested standard CHSH inequality on the Bell state $|\psi\rangle=(|HV\rangle-|VH\rangle)/\sqrt{2}$. 
 Two experiments, one with human-generated random numbers (HRN1) and quantum random numbers (QRN1), the second with real-time human random numbers (HRN2) and the database of humans random numbers (DB2), constitute 4 data sets to analyze.
The random choices $X=0,1$, $Y=0,1$
 define bases of the respective photon measurements of $A=0,1$ and $B=0,1$, i.e. In particular, $XA=00,01,10,11$ correspond to polarization angles 
$0^\circ$, $90^\circ$, $45^\circ$, $135^\circ$ respectively, while $YB=00,01,10,11$ to $-22.5^\circ$, $-112.5^\circ$, $-67.5^\circ$, $22.5^\circ=-157.5^\circ$.
These angles translate to spin correlations (\ref{bellcor}) as $\boldsymbol n=(\cos 2\phi,\sin 2\phi,0)$ so the angles in (\ref{bellcor}) are twice 
the photon polarization angle.
The test assumes no signaling between parties $XA$ and $YB$ but the statistics are postselected on coincidences so we checked it combined with the assumption 
of detection efficiency depending only on the local choice

\begin{table}
\centering{
\begin{tabular}{|c|c|c|c|c|}
\hline
QRN1&00&01&10&11\\
\hline
00& 6260 & 26698& 26678& 4827\\
01& 34992 &5743& 5696&28284 \\
10& 38340& 7610& 6774&30524\\
11& 33318 &6494& 6505&25722\\
\hline
\end{tabular}}
\caption{Full coincidences counts in the QRN1 test for choices $XY$ (rows) and outcomes $AB$ (columns)}\label{qrn1}
\end{table}

\begin{table}
\centering{
\begin{tabular}{|c|c|c|}
\hline
QRN1&00&10\\
\hline
10& 38340& 6774\\
11& 33318 & 6505\\
\hline
\end{tabular}}
\caption{Coincidences counts in the QRN1 test for choices $XY$ (rows) and outcomes $AB$ (columns), for the combined test of ideal angles and input state and  local-dependent detection efficiency
}\label{qrn11}
\end{table}

\begin{table}
\centering{
\begin{tabular}{|C{1cm}|c|c||c|}
\hline
HRN1&0*&1*&$\quad\chi^2\quad$\\
\hline
00& 13775 & 13158 &\\
01& 29096 & 24668 & 63.6~\\
10& 16316 & 13154 &\\
11& 27466 & 22545 & 1.4\\
\hline
\end{tabular}
\begin{tabular}{|C{1cm}|c|c||c|}
\hline
QRN1&0*&1*&$\quad\chi^2\quad$\\
\hline
00& 32958 & 31505 &\\
01& 40735 & 33980 & 159.9\\
10& 45950 & 37298 &\\
11& 39812 & 32227 & 0.07\\
\hline
\end{tabular}
\begin{tabular}{|C{1cm}|c|c||c|}
\hline
HRN2&0*&1*&$\quad\chi^2\quad$\\
\hline
00& 8409 & 7890 &\\
01& 16650 & 14333 & 19.7\\
10& 10333 & 8384 &\\
11& 16069 & 12956 & 0.1\\
\hline
\end{tabular}
\begin{tabular}{|C{1cm}|c|c||c|}
\hline
DB2&0*&1*&$\quad\chi^2\quad$\\
\hline
00& 36403 & 34316 & \\
01& 53112 & 45172 & 108.5\\
10& 42026 & 33546 & \\
11& 51384 & 41504 & 1.4\\
\hline
\end{tabular}}
\caption{Counts $XY$ (rows) for $A=0,1$ (columns) with Bob's outcome ignored (marked by $\ast$) together with Pearson's test of independence of $A$ and $Y$ with given $X$ for each run of experiment 4.}
\label{tt1r}
\end{table}
\begin{table}
\begin{tabular}{|C{1cm}|c|c||c|}
\hline
HRN1&*0&*1&$\quad\chi^2\quad$\\
\hline
00& 13737 & 13196 &\\
10& 15981 & 13489 & 58.6\\
01& 29108 & 24656 &\\
11& 27724 & 22287 & 17.5\\
\hline
\end{tabular}
\begin{tabular}{|C{1cm}|c|c||c|}
\hline
QRN1&*0&*1&$\quad\chi^2\quad$\\
\hline
00& 32938 & 31525 &\\
10& 45114 & 38134 & 139.7\\
01& 40668 & 34027 &\\
11& 39823 & 32216 & 10.01\\
\hline
\end{tabular}
\begin{tabular}{|C{1cm}|c|c||c|}
\hline
HRN2&*0&*1&$\quad\chi^2\quad$\\
\hline
00& 8290 & 8009 &\\
10& 10106 & 8611 & 34.3\\
01& 16638 & 14345 &\\
11& 16091 & 12934 & 18.3\\
\hline
\end{tabular}
\begin{tabular}{|C{1cm}|c|c||c|}
\hline
DB2&*0&*1&$\quad\chi^2\quad$\\
\hline
00& 36028 & 34691 &\\
10& 41066 & 34506 & 168.9\\
01& 53108 & 45176 &\\
11& 51528 & 41360 & 39.8\\
\hline
\end{tabular}
\caption{Counts $XY$ (rows) for $B=0,1$ (columns) with Alice's outcome ignored (marked by $\ast$) together with Pearson's test of independence of $B$ and $X$ with given $Y$ for each run of experiment 4}
\label{tt2r}
\end{table}
\begin{table}
\begin{tabular}{|c|c|c|c|c|}
\hline
total&HRN1&QRN1&HRN2&DB2\\
\hline
00& 26933& 64463 & 16299 & 70719\\
01& 53764& 74695 & 30983 & 98284\\
10& 29470& 83248 & 18717 & 75572\\
11& 50011& 72039 & 29025 & 92888\\
\hline
$\chi^2$& 240.62 & 1561.17 & 228.64 & 312.43\\
\hline
\end{tabular}
\caption{Total counts $XY$ (rows) for each run of experiment 4 together with Pearson's test of independence of $X$ and $Y$.}
\label{tt3r}
\end{table}

In the data, the total counts (Table \ref{tt3r}, first column) depends on the settings. If (\ref{nnpp}) holds then
$N(00)N(11)=N(01)N(10)$ while here e.g. $26933*50011<53764*29470$ beyond statistical error (Pearson's $\chi^2$ independence tests give $\chi^2>240$ so the $p$-value is $<10^{-50}$). Nevertheless, even correcting for the setting-dependent efficiency, the Alice's statistics in Table \ref{tt1r} depends on the Bob's choice, (e.g. $N(0\ast|01)=29096$ is more than twice $N(0\ast|00)=13775$). The Bob's statistics differences (Table \ref{tt2r}) are smaller, e.g. for HRN1 $N(\ast 0|00)=13737$ and 
$N(\ast 0|10)=15981$ give the difference $2244$ at the variance $\sqrt{29718}$ (over $13$ times). At the same time $N(\ast 0|00)=13196$ and $N(\ast 1|10)=13489$
have a difference only $293$ (alternatively, $\chi^2\sim 58.6$ and $p\sim 10^{-13}$).  Note even larger $\chi^2$ for QRN1 and DB2. Trying to compensate the first difference by scaling the efficiencies will increase the second one.
The large deviations are only slightly decreased by look-elsewhere-effect a.k.a. Bonferroni corrections \cite{bonf,lee}, i.e. the $p$-value is increased $4$ times (the number of testable combinations).

%

We see a violation of no-signaling if detection efficiency is only local-choice dependent.
Vice-versa, assuming no-signaling the detection rate must depend on the remote choice. Since only coincidences have been reported, the analysis of signaling could be repeated with the full data, including also single counts. Nevertheless, dropping fair sampling leads also to a reinterpretation of Bell violations, which may disappear in agreement with classical realism based on no-signaling condition.

 The authors claim \cite{scheidl} that a similar apparent violation of the no-signaling assumption has already been observed in previous work by the same group \cite{rauch}, also based on coincidence counts between Alice and Bob. Using single count rates (i.e. not post-selecting on a coincident outcome at the distant location) shows an agreement with no-signaling in that test.
As described in the Supplemental Material of that  publication, the authors show that the effect results from the known efficiency differences of the detectors for outcomes $0$ and $1$ and they verified it with the help of a detailed quantum mechanical model for the experimental results. After correcting the data for the different detection efficiencies, the conditions for no-signaling were fulfilled for the coincidence counts as well. The same polarization analyzer modules, in particular the same single photon detectors, as in \cite{rauch} have been used for the experiment conducted in the course of the BBT collaboration and discussed in the present paper.  

We have checked if the inclusion of different outcome-dependent efficiency can explain one of tests (Table \ref{qrn1}), assuming the standard CHSH test \cite{chsh}
conditions (ideal angles and input state),
i.e. $\tilde{p}(00|01)=\tilde{p}(00|10)$, $\tilde{p}(00|10)=\tilde{p}(00|11)$, 
$\tilde{p}(10|10)=\tilde{p}(10|11)$.
If $p(AB|XY)=\tilde{p}(AB|XY)\eta_a(A)\eta_b(B)$ then  $N(00|01)=N(00|10)$.
Even in the case of detection efficiency depending on both the choice and outcome, e.g. 
\begin{equation}
p(AB|XY)=\tilde{p}(AB|XY)\eta_a(X,A)\eta_b(Y,B),
\end{equation} a subtable \ref{qrn11} should show independence in the sense of $\chi^2$ test while here $\chi^2\sim 28$ ($p\sim 10^{-6}$),
beyond statistical error. It may indicate a drift of general constant non-ideal alignment of measurement angles.

\subsection{5: L. Santodonato et al., Experimental bilocality violation with human randomness}

 The experiment tested local realism with two sources of entangled photons, one sending to $A$ and $B$, the other to $B$ and $C$. Essentially, the setup consists of three observers $A$, $B$, $C$, each measuring $0$ or $1$ at respective choice $X$, $Y$, $Z$. No signaling between each party is assumed. Due to finite detection efficiency, fair sampling is assumed.

We checked if the causal order is satisfied in a sense of no-signaling, i.e. if the probability can depend on the choice not linked causally,
assuming local-choice-dependent efficiencies in the sense of (\ref{nnnppp}). Then the test of (apparent) signaling is the same as in subsection \ref{ex2}.

The summed count for each combination of choices and outcomes is shown in Table \ref{tt1s}.

\begin{table}
\centering{
\begin{tabular}{|c|c|C{0.75cm}|c|c|}
\hline
&000&001&010&011\\
\hline
000& 586& 376& 353& 628 \\
001& 716& 357& 436& 748 \\
010& 1557& 794& 846& 1675 \\
011& 302& 516& 724& 233 \\
100& 741& 548& 551& 844 \\
101& 1395& 939& 1060& 1375 \\
110& 332& 743& 726& 284 \\
111& 763& 172& 208& 513 \\
\hline
\end{tabular}
\begin{tabular}{|c|c|c|c|c|}
\hline
&100&101&110&111\\
\hline
000& 535& 1310& 1220& 569\\
001& 676& 1434& 1476& 611\\
010& 1530& 3225& 3018& 1458\\
011& 1344& 399& 425& 1077\\
100& 704& 1485& 1482& 659\\
101& 1187& 2565& 2586& 1065\\
110& 893& 403& 416& 1010\\
111& 290& 741& 892& 263\\
\hline
\end{tabular}}
\caption{Statistics of coincidences (columns $ABC$) at given choices (rows $XYZ$) of experiment 5}
\label{tt1s}
\end{table}

\begin{table}
\begin{tabular}{|c|c|c|c|c|}
\hline
&00*&01*&10*&11*\\
\hline
000& 962& 981& 1845& 1789\\
001& 1073& 1184& 2110& 2087\\
010& 2351& 2521& 4755& 4476\\
011& 818& 957& 1743& 1502\\
100& 1289& 1395& 2189& 2141\\
101& 2334& 2435& 3752& 3651\\
110& 1075& 1010& 1296& 1426\\
111& 935& 721& 1031& 1155\\
\hline
\end{tabular}
\begin{tabular}{|c|c|c|c|c|}
\hline
&0*0&0*1&1*0&1*1\\
\hline
000& 939 & 1004 & 1755 & 1879\\
001& 1152 & 1105 & 2152 & 2045\\
010& 2403 & 2469 & 4548 & 4683\\
011& 1026 & 749 & 1769 & 1476\\
100& 1292 & 1392 & 2186 & 2144\\
101& 2455 & 2314 & 3773 & 3630\\
110& 1058 & 1027 & 1309 & 1413\\
111& 971 & 685 & 1182 & 1004\\
\hline
\end{tabular}
\begin{tabular}{|c|c|c|c|c|}
\hline
&*00&*01&*10&*11\\
\hline
000& 1121 & 1686 & 1573 & 1197\\
001& 1392 & 1791 & 1912 & 1359\\
010& 3087& 4019& 3864 & 3133\\
011& 1646 & 915 & 1149 & 1310\\
100& 1445 & 2033 & 2033 & 1503\\
101& 2582 & 3504 & 3646 & 2440\\
110& 1225 & 1146 & 1142 & 1294\\
111& 1053 & 913 & 1100 & 776\\
\hline
\end{tabular}
\caption{Counts of Table \ref{tt1s} with one of outcomes ignored (marked by $\ast$)}
\label{tt2s}
\end{table}
\begin{table}
\centering{
\begin{tabular}{|c|c|c|}
\hline
&0**&1**\\
\hline
000& 1943 & 3634\\
001& 2257 & 4197\\
010& 4872 & 9231\\
011& 1775 & 3245\\
100& 2684 & 4330\\
101& 4769 & 7403\\
110& 2085 & 2722 \\
111& 1656 & 2186\\
\hline
\end{tabular}
\begin{tabular}{|c|c|c|}
\hline
&*0*&*1*\\
\hline
000& 2807 & 2770\\
001& 3183 & 3271\\
010& 7106 & 6997\\
011& 2561 & 2459\\
100& 3478 & 3536\\
101& 6086 & 6086\\
110& 2371 & 2436\\
111& 1966 & 1876\\
\hline
\end{tabular}
\begin{tabular}{|c|c|c|c|}
\hline
&**0&**1 & total\\
\hline
000& 2694 & 2883 & 5577\\
001& 3304 & 3150 & 6454\\
010& 6951 & 7152 & 14103\\
011& 2795 & 2225 & 5020\\
100& 3478 & 3536 & 7014\\
101& 6228 & 5944 & 12172\\
110& 2367 & 2440 & 4807\\
111& 2153 & 1689 & 3842\\
\hline
\end{tabular}}
\caption{Counts of Table \ref{tt1s} with two outcomes ignored (marked by $\ast$) and total numbers of coincidences}
\label{tt3s}
\end{table}
\begin{table}
\centering{
\begin{tabular}{|c|c|c|c|c|}
\hline
& 00 & 01 & 10 & 11\\
\hline
X and A& 15.69 & 31.81 & 120.15 & 55.007 \\
Y and B& 0.004 & 3.2 & 0.07 & 1.60 \\
Z and C& 9.97 & 60.48 & 4.44 & 39.55 \\
\hline
\end{tabular}}
\caption{$\chi^2$ values calculated for each party. Each test is performed for other choices fixed; their values are given in the upper row in alphabetic order (XY, YZ or XZ).}
\label{tt4s}
\end{table}
\begin{table}
\centering{
\begin{tabular}{|c|c|c|c|c|}
\hline
& 00 & 01 & 10 & 11\\
\hline
X and B& 0.69 & 0.78 & 1.61 & 0.02 \\
X and C& 2.04 & 0.001 & 0.003 & 0.11 \\
Y and A& 0.15 & 0.18 & 30.92 & 18.70 \\
Y and C& 1.54 & 22.80 & 0.13 & 27.78 \\
Z and A& 0.02 & 1.07 & 1.56 & 0.06 \\
Z and B& 1.22 & 0.58 & 0.30 & 2.91 \\
\hline
\end{tabular}}
\caption{$\chi^2$ values calculated for independent choices and outcomes, as assumed. Each test is performed for other choices fixed; their values are given in the upper row in alphabetic order (XY, YZ or XZ).}
\label{tt4sa}
\end{table}

In the data, in the last of Tables \ref{tt3s}  that total counts still depend on the choices but not independently i.e. the assumption (\ref{nnnppp}) implies $N(010)N(101)=N(000)N(111)$ while $14103*12172\neq 5577*3842$ beyond statistical error. Therefore it may indicate the fail of the assumption (\ref{nnnppp}), not no-signaling. In view of combined choice-dependent efficiency we 
cannot directly check no-signaling (which apparently would be immediately violated) but rather no-signaling combined with independent efficiency.
So we test independence, e.g. if $p(\ast 00|010)p(\ast 01|110)=p(\ast 01|010)p(\ast 00|110)$ by Pearson's $\chi^2$ test. Here in Table \ref{tt2s} column 1 and 2, rows 3 and 7, $\chi^2\simeq 48$ giving $p$-value of the order $10^{-11}$.
It is only slightly decreased by look-elsewhere-effect, i.e. the $p$-value is increased about $100$ times (the number of testable combinations). Another example is $p(0\ast\ast|100)p(1\ast\ast|111)=p(1\ast\ast|100)p(0\ast\ast|111)$ 
giving (Table\ref{tt4sa}) $\chi^2\simeq 24$ or $p$-values of the order $10^{-5}$. Nevertheless most combination give much smaller $\chi^2$, often with $p\sim 1$.

The large deviations indicate that
it can be either no-signaling or local choice dependent efficiency assumption to fail. Since only coincidences have been reported, the analysis of signaling could be repeated with the full data, including also single counts. According to authors \cite{scia} the apparent violation may be caused by the long runtime of the experiment when the efficiencies undergo systematic time-dependent bias.

\subsection{6: K. Redeker, R. Garthoff, D. Burchardt, H. Weinfurter, W, Rosenfeld, Violation of Bell\'s inequality with a single atom and single photon entangled over a distance of 400m}
 
 The experiment tested standard CHSH inequality on the entangled photon-atom Bell state $|\psi\rangle=(|\downarrow L\rangle+|\uparrow R\rangle)/\sqrt{2}$. 
 in the $LR$ photon ($A$) basis and $\uparrow\downarrow$ atom ($B$) basis.
The measurement of the photon by avalanche photodiodes (APD) depended on the passive choice $X=0,1$ while the atom measurement (by ionization) depended on the choice $Y=0,1$,
realized by either a quantum random number generator (QRN) or human random number generator (HRN). The dichotomic outcomes of the respective measurements, $A=0,1$ and $B=0,1$ depend on the choices ($X$ and $Y$, respectively). In particular, $XA=00,01,10,11$ correspond to polarization angles 
$0^\circ APD2$, $0^\circ APD4$, $45^\circ APD1$, $45^\circ APD3$, while $YB=00,01,10,11$ to $22.5^\circ\mathrm{in}$, $22.5^\circ\mathrm{out}$, $-22.5^\circ\mathrm{in}$, $-22.5^\circ\mathrm{out}$. The angles enter the Bell correlation formula (\ref{bellcor}) multiplied by 2, as in subsection \ref{sec4}.
 The test assumes no signaling between parties $XA$ and $YB$ but also fair sampling at the side of the photon ($A$).

We checked  no-signaling assuming equal APD efficiencies i.e. if $p(\ast B|0Y)=p(\ast B|1Y)$ and $p(A\ast|X0)=p(A\ast|X1)$ for all runs.
In the first case, one has to take into account fair sampling and passive choice $X$ (the signaling could be in principle subluminal).
We also checked if nonequal efficiencies explain the data under that assumption of the ideal Bell state and measurement angles.

\begin{table}
\centering{
\begin{tabular}{|c|c|c|c||c|}
\hline
HRN&0*&1*&**&$\chi^2$\\
\hline
00& 2883 & 2462 &5345 & \\
01& 2635 & 2253 & 4888 & $10^{-3}$\\
10& 2697 & 2224 & 4921 &\\
11& 2520 & 2041 & 4562 & 0.19\\
\hline
\end{tabular}
\begin{tabular}{|c|c|c|c||c|}
\hline
QRN&0*&1*&**&$\chi^2$\\
\hline
00& 2845 & 2403 & 5248 &\\
01& 2773 & 2335 & 5108 & 0.006\\
10& 2644 & 2137 & 4781 &\\
11& 2658 & 2103 & 4761 & 0.27\\
\hline
\end{tabular}}
\caption{Counts $XY$ (rows) for $A=0,1$ (columns) with Bob's outcome ignored (marked by $\ast$), and the total counts, together with Pearson's test of independence of $A$ and $Y$ with given $X$ for each run of experiment 6}
\label{tt1u}
\end{table}
\begin{table}
\centering{
\begin{tabular}{|c|c|c||c|}
\hline
HRN&*0&*1&$\chi^2$\\
\hline
00& 2707 & 2638 &\\
10& 2440 & 2481 & 1.16\\
01& 2285 & 2603 &\\
11& 2470 & 2092 & 51\\
\hline
\end{tabular}
\begin{tabular}{|c|c|c||c|}
\hline
QRN&*0&*1&$\chi^2$\\
\hline
00& 2649 &2599 &\\
10& 2368 & 2413 & 0.9\\
01& 2387 & 2721 & \\
11& 2648 & 2113 & 78\\
\hline
\end{tabular}}
\caption{Counts $XY$ (rows) for $B=0,1$ (columns) with Alice's outcome ignored (marked by $\ast$) together with Pearson's test of independence of $B$ and $X$ with given $Y$ for each run of experiment 6}
\label{tt2u}
\end{table}


In the data, the total counts (Table \ref{tt1u}) depends on the settings.
In the  HRN run, there is additional dependence of the total counts on the atom setting but uncorrelated ($\chi^2\simeq 0.23$). 
The reason might be asymmetry in human choice.
As regards no-signaling, in the case $Y\to A$ it cannot be rejected. However, dependence of Bob's outcome $B$ on the photon setting $X$, under the assumption of equal detection efficiencies, is clear in the $\chi^2$ test, in both runs, i.e. the independence test $p(\ast 0|01)p(\ast 1|11)=p(\ast 0|11)p(\ast 1|01)$ fails ($\chi^2\simeq 51$ for HRN, $p\sim 10^{-11}$  and $78$ for QRN, $p\sim 10^{-17}$ ). We find that
$N(\ast 1|01)N(\ast 0|11)/N(\ast 1|11)N(\ast 0|01)$ is $1.34$ for HRN while $1.42$ for QRN, which is moderately consistent. The authors confirmed that the efficiencies were not equal \cite{kai}.
We have also checked if this effect could be explained by different efficiencies of APD, assuming correct input state and detectors' angles. Suppose the APD$n$ has efficiency $\eta_n$, for $n=1,2,3,4$.
If the measurement axes are as in standard CHSH test \cite{chsh}), then the probabilities is as in Table \ref{tt4u}. Then still $\chi^2$ for QRN in the first two rows of (\ref{tt3u}) gives $\chi^2\sim 14$ ($p\sim 2\cdot 10^{-4}$) while the lower two rows (in case of alternative configuration) gives  $\chi^2\sim 35$ ($p\sim 10^{-8}$). Therefore this explanation seems insufficient. Possible further reasons may include unspecified deviation from the ideal Bell state or measurement axes. The $p$-value is increased $4$ times by the look-elsewhere-effect.

\begin{table}
\centering{
\begin{tabular}{|c|c|c|}
\hline
&*0&*1\\
\hline
10&$\eta_1s_++\eta_3s_-$&$\eta_1s_-+\eta_3s_+$\\
11&$\eta_1s_++\eta_3s_-$&$\eta_1s_-+\eta_3s_+$\\
\hline
00&$\eta_2s_++\eta_4s_-$&$\eta_2s_-+\eta_4s_+$\\
01&$\eta_2s_++\eta_4s_-$&$\eta_2s_-+\eta_4s_+$\\
\hline
\end{tabular}}
\caption{Influence of asymmetric detection efficiencies on detection probabilities, with $4s_\pm =1\pm 1/\sqrt{2}$ in the ideal Bell test in the case angles
as in Sec. \ref{sec4} -- upper two rows and with Alice's angles additionally rotated by $45^\circ$} -- lower two rows
\label{tt3u}
\end{table}

\begin{table}
\centering{
\begin{tabular}{|c|c|c||c|}
\hline
HRN&*0&*1&$\chi^2$\\
\hline
00& 2707 & 2638 & \\
01& 2285 & 2603 & 15.5\\
10& 2440 & 2481 & \\
11& 2470 & 2092 & 19.7\\
\hline
\end{tabular}
\begin{tabular}{|c|c|c||c|}
\hline
QRN&*0&*1&$\chi^2$\\
\hline
00& 2649 &2599 &\\
01& 2387 & 2721 & 14.5 \\
10& 2368 & 2413 & \\
11& 2648 & 2113 & 35.5\\
\hline
\end{tabular}}
\caption{In the case of ideal Bell angles, the outcome  $B$ should not depend on $Y$ if either $X=1$ or $X=1$ for each run of experiment 6}
\label{tt4u}
\end{table}

The different total counts can be explained by different APD detection efficiencies but it should be confirmed
by a diagnostic run.  In the HRN test, there is human choice asymmetry, similar to experiment 13 (see later). 
On the other hand, the correlation between the photon's (A) setting and the atom's (B) outcome requires either some additional signal from the source to both parties or different efficiencies but combined with nonideal Bell state and/or measurement axes. In any case, in future experiments, these effects should be independently identified by diagnostic runs, preferably before the main test.

\subsection{9: P. Farrera, G. Heinze, H. de Riedmatten, Bell test using entanglement between a photon and a collective atomic excitation, driven by human
randomness}
 
 The experiment tested standard CHSH inequality on the entangled atom-photon Bell state which can be translated into photon-photon state $|\psi\rangle=(|E_wE_r\rangle+e^{i\phi}|L_wL_r\rangle)/\sqrt{2}$ ($E,L$ stand for early,late, $w,r$ -- write,read, $\phi$ phase to adjust.
 The choices and measurement can be translated into Alice-Bob $0,1$ numbers as follows.
$A=0,1$ or $B=0,1$ corresponding to Alice's and Bob's $+,-$ outcomes, while $X=0,1$ correspond to Alice's choice $w,w'$ and $Y=0,1$ -- to $r,r'$
The measurement of the photon depended on the passive choice $X=0,1$ while the atom measurement dependent on the choice $Y=0,1$,
 It assumes no signaling between parties $AX$ and $BY$ but also fair sampling, due to low detection efficiency, $\sim 5\%$.
We checked no-signaling, assuming local choice dependent efficiency (\ref{nnpp}) i.e. if $\tilde{p}(\ast B|0Y)=\tilde{p}(\ast B|1Y)$ and $\tilde{p}(A\ast|X0)=\tilde{p}(A\ast|X1)$ for all runs.
We also checked if the rate of coincidences/trials depends on choices.

\begin{table}
\centering{
\begin{tabular}{|c|c|c|c|c|}
\hline
&0*&1*&**&trials\\
\hline
00& 65 & 124 & 189&  94433\\
01& 129 & 213 & 342 & 191674\\
10& 81 & 275 & 356 &   191332\\
11& 63 & 150 & 213 & 122561\\
\hline
\end{tabular}
\begin{tabular}{|c|c|c|}
\hline
&*0&*1\\
\hline
00& 59 & 130\\
01& 166 & 176\\
10& 102 & 254\\
11& 145 & 68\\
\hline
\end{tabular}}
\caption{Counts at choices $XY$ (rows) for $B=0,1$ (columns) with Alice's or Bob's outcome ignored (marked by $\ast$) for each run, and total coincidences and trials in experiment 9}
\label{tt2z}
\end{table}
\begin{table}
\centering{
\begin{tabular}{|c|c|}
\hline
X& $\chi^2$ \\
\hline
0& 3.28 \\
1& 0.58 \\
\hline
\end{tabular}
\begin{tabular}{|c|c|}
\hline
Y& $\chi^2$ \\
\hline
0& 20.34 \\
1& 0.39 \\
\hline
\end{tabular}}
\caption{Pearson's test of independence of $A$ and $Y$ (left) and of $B$ and $X$ (right) for experiment 9}
\label{ttz_chi}
\end{table}


In the data, the total coincidence counts (Table \ref{tt2z}) are roughly proportional to the number of trials but the latter depends on both settings
in a correlated way (i.e. Alice and Bob do not choose the settings independently). No-signaling is violated as  $\chi^2\sim 20$ ($p\sim 10^{-5}$) for (\ref{tt2z}) taking the second and fourth row of $\ast 0$ and $\ast 1$ but it may be explained alternatively by correlation-dependent detection efficiency \cite{pau2}.
The $p$-value is increased $4$ times by the look-elsewhere effect.

The observed deviations put in question the local dependency of efficiency but may also indicate some form of signaling. In future experiments, these effects should be independently identified by diagnostic runs, preferably before the main test.

\subsection{10: A. Lenhard, A. Seri, D. Riel\"ander, O. Jimenez, A. Mattar, D. Cavalcanti, M.
Mazzera, A. Acin, and H. de Riedmatten, Violation of a Bell inequality using high-dimensional frequency-bin entangled photons}

 The experiment tested Clauser-Horne (CH) inequality \cite{chineq} on the entangled two-photon Bell state, one idler (Alice $A$) and  signal (Bob $B$).
 The measurement can be translated into Alice $0,+1,-1$ (or nothing) by taking three different frequency modes. On the other hand, Bob always measures $0$ or nothing (missing photon due to low detection efficiency). Alice can choose $X=0,1$ (by changing modulation depth and phase) while Bob
can choose $Y=nm$ where $n=0,1$ (modulation depth) while $m=1,..,16$ (phase). The CH inequality is measured assuming fair sampling and
symmetry of correlations. Bob's labels $4$ and $14$ correspond to real phases $170$ and $342$, respectively, and were used to find Bell inequality violation.

Since Bob's outcome was always $0$ or nothing, the most appropriate test is one-sided signaling, i.e. if Bob's probabilities equality
 $p(\ast 0|0Y)=p(\ast 0|1Y)$ hold.
 We checked no-signaling, assuming local choice dependent efficiency (\ref{nnpp}) i.e. if $\tilde{p}(\ast 0|0Y)=\tilde{p}(\ast 0|1Y)$.

\begin{table}
\begin{tabular}{|c|r|r|}
\hline
ph &00&01\\
\hline
140&	7.83$\pm$0.56&	6.17$\pm$0.54\\
151&	10.67$\pm$0.74&	8.75$\pm$0.83\\
160&	7.00$\pm$0.92&	6.96$\pm$0.79\\
170&	8.98$\pm$0.73&	6.78$\pm$0.75\\
183&	9.00$\pm$0.75&	7.03$\pm$0.64\\
197&	8.91$\pm$0.83&	7.45$\pm$0.96\\
215&	8.82$\pm$0.85&	9.48$\pm$1.34\\
232&	8.17$\pm$0.61&	6.99$\pm$0.76\\
250&	5.79$\pm$0.66&	8.24$\pm$0.76\\
267&	7.04$\pm$0.78&	7.37$\pm$0.87\\
284&	7.14$\pm$0.77&	7.28$\pm$0.75\\
298&	6.70$\pm$0.79&	7.05$\pm$0.87\\
321&	7.82$\pm$0.76&	5.43$\pm$0.76\\
342&	6.01$\pm$0.65&	5.71$\pm$0.77\\
352&	7.07$\pm$0.67&	6.99$\pm$0.82\\
358&	7.91$\pm$0.63&	4.56$\pm$0.58\\
all&	124.87$\pm$2.95& 112.25$\pm$3.27\\
\hline

\end{tabular}
\begin{tabular}{|c|r|r|c|}
\hline
ph &10&11& $\chi^2$\\
\hline
140&5.44$\pm$0.47&	4.20$\pm$0.42&	0.01\\
151&6.17$\pm$0.73&	3.15$\pm$0.48&	4.80\\
160&4.48$\pm$0.77&	4.08$\pm$0.68&	0.09\\
170&6.60$\pm$0.73&	3.39$\pm$0.50&	2.97\\
183&6.20$\pm$0.77&	4.33$\pm$0.63&	0.26\\
197&5.18$\pm$0.68&	3.80$\pm$0.59&	0.27\\
215&6.03$\pm$0.72&	3.94$\pm$0.69&	3.78\\
232&5.80$\pm$0.77&	5.00$\pm$0.65&	0.00\\
250&6.92$\pm$0.63&	5.71$\pm$0.64&	7.23\\
267&7.54$\pm$0.82&	2.82$\pm$0.67&	13.98\\
284&5.44$\pm$0.71&	5.61$\pm$0.82&	0.00\\
298&5.55$\pm$0.69&	6.40$\pm$0.82&	0.14\\
321&6.84$\pm$1.03&	4.87$\pm$0.60&	0.01\\
342&7.97$\pm$0.86&	6.64$\pm$0.77&	0.32\\
352&5.90$\pm$0.63&	7.13$\pm$0.83&	0.89\\
358&7.98$\pm$0.74&	7.51$\pm$0.74&	6.18\\
all&100.06$\pm$2.97& 78.59$\pm$2.67&	5.37\\
\hline

\end{tabular}

\caption{Weighted counts $C/T$ and errors $e^2=C/T^2$ (at $\pm$ sign) at choices $XY$ (columns) for $B=0$ (columns) with Alice's outcome ignored for special phases (ph -- rows), or all summed in experiment 10 and $\chi^2$ in the last column}
\label{tt2y}
\end{table}


To compare the counts they must be weighted by the number of trials (10s intervals, when the coincidences for a particular choices are collected)
as shown in the Table \ref{tt2y}. The weighted trials $N$ are sums of $C/T$ while $e^2$ is the sum of $C/T^2$ for $C$ coincidences and $T$ trials.
We adjusted  $\chi^2$ test formula for varying trials with 
\begin{equation}
\chi^2=\left(\frac{N_{00}N_{11}-N_{01}N_{10}}{N_{00}+N_{01}+N_{10}+N_{11}}\right)^2\sum_{ij}e^{-2}_{ij}
\end{equation}
 The numbers of trials $T$ vary between 2 and 36.
For the phases used in CH test, 170 and 342 but also 160, no-signaling is consistent with $\chi^2$ equal $2.97$, $0.32$, and $0.09$, respectively.
In the data, there is an apparent violation of no-signaling or violation of local-dependent efficiency (\ref{nnpp}) at phase 267 of moderate significance where $\chi^2\sim 14$
giving $p$-value $\sim 2\cdot 10^{-4}$  increased $16$ times (number of phases) by the look-elsewhere-effect.
It may also indicate some form of signaling but also nonlocal-dependent efficiency.
According to the authors \cite{acav} some problems could have occurred during the measurement for some point (e.g. laser temporarily out of a lock, leading to reduced coincidences for this setting and phase). The point at phase 267 is clearly below the curve in Supplementary Fig. 15 of BIG Bell test \cite{bbt}. It might be also a small difference in fiber coupling for the different settings, which  could lead to a difference in efficiency  (locally for Alice and Bob) for the different settings. In future experiments, these effects should be independently identified by diagnostic runs, preferably before the main test.

\subsection{12: J. Cari\~ne et al. Post-selection loophole-free energy-time Bell test fed with human-generated inputs}

The authors report an optical experiment employing Franson "hug" configuration -- a variation that avoids the post-selection of results present in basic Franson configuration. In the "hug" configuration, if the two emitted photons are detected by Alice and Bob then it is guaranteed that they both traveled short ways ($S$) or long ways while obtaining phase shifts $\phi_A$ and $\phi_B$ ($L$). The results when single photons are measured by either Alice or Bob are discarded, due to low detection efficiency. Thus fair sampling of a sort is inherent here. The remaining could be assumed to have been in a state $|\psi\rangle=\frac{1}{\sqrt{2}}(|SS\rangle+e^{i(\phi_a+\phi_b)}|LL\rangle)$. CHSH inequality was tested, with varying phases set by human-generated input. We assume the fair sampling for the empty counts is also applied. As communicated, the closing of detection loophole was not the purpose of the experiment \cite{xavier}. The setting at Alice and Bob are given by $X$ and $Y$, respectively, equal $0$ or $1$, corresponding to $\phi_a=\pi/4$, $\phi_b=0$ or
$\phi'_a=-\pi/4$, $\phi'_b=\pi/2$. The outcomes $A,B$ were also $0$ or $1$ (or nothing).

Based on the data provided and assuming local dependent efficiency (\ref{nnpp}), we assessed the number of empty counts and consequently the efficiency of the detectors. Further, we checked no-signaling i.e. if $\tilde{p}(\ast B|0Y)=\tilde{p}(\ast B|1Y)$ and $\tilde{p}(A\ast|X0)=\tilde{p}(A\ast|Y0)$. We also checked if the data are consistent with equal and independent detection efficiency, i.e. if the probability $p(AB|XY)$ is consistent with the single efficiency $\eta$ and Bell state and angles $\phi^{(\prime)}_{a,b}$.
For arbitrary phases the correlations can be written in the form  similar to (\ref{bellcor}), $\langle AB\rangle=E_{ab}=v\cos(\phi_a-\phi_b)$ with the visibility $v$. In the ideal Bell test
$E_{ab}=E_{ab'}=E_{a'b}=-E_{a'b'}$.

The total number of coincidences depends on settings but independently, with $\chi^2\sim 0.3$ for the last column of Table \ref{tt2x}.
Taking this effect into account, there is no signaling observed within the standard error.
However, the correlations do not agree with the ideal Bell model, eg. $E_{a'b}$ is not equal $E_{a,b'}$ i.e.
It can be quantified defining $M=NE$ where $N$ is the total number of coincidences. Here $M$ is calculated using Table \ref{tt1x} while $N$ is in Table \ref{tt2x}. The $\chi^2$ test for $M_{a'b}$, $M_{ab'}$, $N_{a'b}$, $N_{ab'}$
gives $\chi^2\sim 348$ ($p<10^{-70}$). The difference can be however explained by a small long-term  uncontrolled phase drift, confirmed by the authors \cite{xavier}.
Taking into account setting-dependent detection efficiency, no signaling signatures have been found. However, the strong deviation from the ideal Bell model should be confirmed in a diagnostic run, testing solely phase modulators.

\begin{table}
\centering{
\begin{tabular}{|c|c|c|c|c|}
\hline
&00&10&01&11\\
\hline
00& 2296 & 647 & 672 & 2248 \\
01& 1796 & 762 & 717 & 1781 \\
10& 2332 & 324 & 282 & 2338 \\
11& 318 & 1908 & 1896 & 359 \\
\hline
\end{tabular}}
\caption{Total counts at choices $XY$ (rows) for each configuration $AB$ (columns)}
\label{tt1x}
\end{table}

\begin{table}
\centering{
\begin{tabular}{|c|c|c|}
\hline
&0*&1*\\
\hline
00& 2968 &2895\\
01& 2513 &2543\\
10& 2614 & 2662 \\
11& 2214 & 2267 \\
\hline
\end{tabular}
\begin{tabular}{|c|c|}
\hline
*0&*1\\
\hline
2943 & 2920\\
2558 & 2498\\
2656 & 2620\\
2226 & 2255\\
\hline
\end{tabular}
\begin{tabular}{|c|}
\hline
total\\
\hline
5863 \\
5056 \\
5276 \\
4481 \\
\hline
\end{tabular}}
\caption{Counts at choices $XY$ (rows) for $A=$0,1 (columns) with Bob's outcome ignored (marked by $\ast$) on the left and for $B=$0,1 with Alice's outcome ignored in the middle and a total count on the right in experiment 12}\label{tt2x}
\end{table}

\subsection{13: L.K. Shalm et al., Using human generated randomness to violate a Bell inequality without detection or locality loopholes}

The experiment is the standard two-party (Alice and Bob) Bell test, with two entangled (not maximally) photons detected at optimal polarization angles
chosen by $X=0,1$ by Alice and $Y=0,1$ by Bob.
Due to imperfect detection efficiency, the configuration differs from the ideal CHSH model. To increase statistics, each of observers can detect a photon in one of $16$ time bins. Nevertheless, it allows to test if the correlations violate local realism, i.e. existence of the joint probability of outcomes depending on the local choice. The influence of the remote choice is excluded by relativity, i.e. the time between the choice and the end of the measurement is shorter than the light-speed signal. 

We checked no-signaling i.e. if $p(\ast B|0Y)=p(\ast B|1Y)$ and $p(A\ast|X0)=p(A\ast|Y0)$ where $A,B=0,..,16$ are the photon detections in the appropriate time bin, with $0$ standing for no detection. We have taken into account the fact noted by the authors that $X,Y=0$ are chosen about $5\%$ more often than $1$.
As a stopping criterion, we took the last $XY=11$ event.

Taking the last column $\ast 0$ or $0\ast$ as a reference we find in the independence test
$\chi^2\sim 8$ giving $p\sim 0.005$ for $A=11$ and $X=0$. Correcting by the look-elsewhere effect this is increased $32$ giving $p\sim 0.16$, consistent with no-signaling assumption. It is also clear that there is a bias in the settings choices probabilities. The bias is independent ($\chi^2\sim 1$) but different for the two parties ($>60$ variances). It is also evident (but less significant) with the original authors' stopping criterion giving
$N_{10}=10125716$ and $N_{01}=10105777$ ($\sim 4.43$ variances) for the human test.
Within the available data, no signaling signatures have been found. However, the setting choice bias is asymmetric.

\begin{table}
\begin{tabular}{|C{0.45cm}|cccc|}
\hline
&00&10&01&11\\
\hline
1*&2195&7185&1935&6088\\
2*&2233&7695&1944&6675\\
3*&2119&7672&2007&6739\\
4*&2136&7652&1997&6841\\
5*&2310&7692&1967&6708\\
6*&2228&7765&1934&6798\\
7*&2205&7690&1956&6908\\
8*&2206&7927&1921&6931\\
9*&2195&7953&1941&6824\\
10*&2299&7995&1905&6788\\
11*&2320&7900&1869&7016\\
12*&2245&7886&1931&6911\\
13*&2241&7755&2093&7124\\
14*&2234&7831&2013&6921\\
15*&2169&7856&2009&6958\\
0*&11636736&10510154&10237027&9240316\\
\hline
\end{tabular}

\begin{tabular}{|C{0.45cm}|cccc|}
\hline
&00&10&01&11\\
\hline
*1&2197&1945&6651&6138\\
*2&2322&2014&6491&5860\\
*3&2217&1991&6912&6214\\
*4&2186&2014&7108&6291\\
*5&2219&1996&7155&6429\\
*6&2249&1964&6881&6396\\
*7&2194&1977&7051&6434\\
*8&2218&2055&7055&6485\\
*9&2222&2085&7221&6473\\
*10&2239&2061&7210&6677\\
*11&2346&2003&7195&6507\\
*12&2251&2045&7238&6551\\
*13&2172&1991&7437&6631\\
*14&2245&2022&7429&6619\\
*15&2181&2092&7349&6713\\
*0&11636613&10596353&10160066&9246131\\
**&11670071&10626608&10266449&9342546\\
\hline
\end{tabular}

\caption{Counts at choices $XY$ (columns) with either Bob's or Alice's outcome ignored (rows $AB$ with the mark $\ast$ in experiment 13) }\label{tt2w}
\end{table}

\section{Discussion}

In the data received from  BBT experiments, there are observed disagreements with various simple assumptions.
Firstly, the human choice is not perfectly random but biased with more $0$s than $1$s (experiments 6 and 13).
It is also highly anticorrelated, with very likely sequences $0101$ and $1010$ (experiment 1).
The data from other experiments depend on too many parameters to confirm these observations.
Nevertheless, a biased human choice can be simply incorporated into $\eta$ in (\ref{nnpp}), so it does not affect our analysis based on independence tests.
Secondly, the photon detection efficiency is often different for different detectors (experiment 6, communicated by the authors \cite{kai}, also mentioned as possible by the authors of experiment 10 \cite{acav}) and dependent on the state of the other part of the setup (experiment 2). Thirdly, the actual quantum state and measurement axes are different from the ideal Bell case (experiment 6 and 12), probably by some phase drifts (as suggested in the case of 12 by the authors \cite{xavier}) or misalignments. Ignoring these effects could lead to apparent signaling, also in experiments 4,5,9,10. 

These observations suggest for future tests of local realism (a) to narrow the problem of varying detection efficiency e.g. by its better control, (b) to block known communication between detectors, (c) to collect sufficient data to check no-signaling, preferably with closed loopholes,
 (d) to focus more attention on diagnostic runs checking the input state, operation flow, and detectors. The last suggestion follows also from the general expectation from quantum tests of local realism - they are not only performed to violate local realism but also to confirm quantum predictions. 
In the future, one can also test if the same time-tagged stream of choices give the same statistical results for different experiments in separate locations.
Due to the lack of global synchronization between experiments, we were unable to test it.

\section*{Acknowledgements}
We thank M. Mitchell, R. Patel, M. Ringbauer, T. Scheidl, F. Sciarrino, R. Chavez, K. Redeker, W. Rosenfeld, H. de Riedmatten, P. Farrera, A. Seri, G.B. Xavier, L.K. Shalm, A. Acin, D. Cavalcanti for making available data from the experiments, helpful explanations, and comments.

\end{document}